# The Impact of Public Safety Measures on the Spread of COVID-19 in the United States Assessed By Causal Model-Based Projections of the Pandemic

Keshav Amla[1]* & Tarun Amla[2]*

[1] *Avishtech, Inc., 1798 Technology Dr., Suite 1794, San Jose, CA, USA, 95110*
[2] *School for Engineering of Matter, Transport and Energy, Arizona State University, Tempe, AZ, USA 85287*
* These authors contributed to this work equally. The names appear in alphabetical order.

Corresponding Author: Keshav Amla, Avishtech, Inc. (keshav.amla@avishtech.com)

The novel coronavirus, SARS-CoV-2, and the disease it causes, COVID-19 was declared a pandemic on March 11, 2020 by the World Health Organization. Since then, the disease has spread all over the world, with the United States becoming the country with the highest number of cases. Governments around the world have undertaken varying degrees of public safety measures, including recommendations and ad campaigns for improved hygiene practices, enacting social distancing requirements and limiting large public gatherings, and stay-at-home orders and lockdowns. In the United States, while the response has varied greatly from state to state, we clearly see that the effect of these public safety measures is significant and, if these measures continue to remain in effect, or are expanded to a nationwide lockdown, the pandemic can be controlled and the disease likely overcome with mitigated consequences. In this paper, we model the spread of the novel coronavirus using a causal model. We find that, with continued lockdown measures, the United States can, according to this model, limit the total number of infections to ~1.35 million and the total number of deaths to ~72 thousand. A 60 day lockdown can save countless lives.

## I. Introduction and Motivation

COVID-19 was first observed as a zoonotic disease in a human in December of 2019 in Wuhan. Because SARS-CoV-2, the virus that confers COVID-19, has several methods of spread and because of its relative ease of communicability from human to human, this disease has been spreading around the world at a very high pace, having been declared a global pandemic by the World Health Organization on March 11, 2020.

Countries all around the world have been working to try to combat the pandemic through public safety measures, which predominantly involve promoting improved hygiene practices, like frequent and lengthy hand washing, enactment of social distancing measures, such as guidelines for keeping more than six feet apart from others and cancellation of large events and bans on gatherings of more than a few people, and, stay-at-home or lockdown orders, requiring the closing of non-essential businesses and banning outdoor movement, except as it pertains to health, sustenance necessities, or those businesses deemed essential. This is in addition to schools and universities having been closed, and most workplaces shifting exclusively to a work-from-home model.

The United States has become the worst hit in terms of the number of cases, and, as a result, has enacted many of these public safety measures. Though there is no nationwide lockdown, most states have issued stay-at-home orders and have closed all non-essential businesses, in addition to social distancing measures.

The current social distancing and lockdown measures are vital to our society's survival and we simply wished to provide a brief, simple analysis to make it clear why that is and to implore people to obey these measures. We further argue that increased social distancing measures can serve to further prevent this disease from getting completely out of control.

We are not epidemiologists, but rather engineers. The bulk of what is being pushed by experts in the field, however, is proving inaccurate due to the reliance on analogy to previous outbreaks of similar viruses, and also ineffective in making the case for the importance of continued implementation and obedience of these public health measures. As a result, we intend to make this case in a clear, accessible fashion in this short paper.

In this paper, we present a statistical treatment of the COVID-19 outbreak in the United States. This is implemented through the use of a causal model to describe the effect of the public safety measures as enacted. This causal model is used to describe the decreased growth rate, in the days since strong public safety measures have gone into effect in the country. This model allows us to make reasonable, indicative projections on the spread of the disease and its impact in terms of hospitalizations and deaths over time.

The paper is organized as follows: in Section II, we introduce and justify our assumptions and methodology for building our model, which is constructed using the data available from the COVID Tracking Project [1]. Next, in Section III, we show the results of the projection of our model forward in time for infections, hospitalizations, and deaths, and highlight the impact of the public safety measures. Lastly, we provide a discussion of the primary takeaways of our findings and conclude with some recommendations, as well as acknowledge the limitations of the analysis conducted herein.

## II. Methodology

The limited understanding of the virus epidemiologically has limited the ability of experts to apply science-derived models that depend on knowing several key properties of the virus and the kinetics and means of its spread, its incubation period, and additional mechanistic behavior about SARS-CoV-2. While we recognize that these types of science-based models (which may or may not be enhanced by statistical learning) are generally very powerful and tend to prove superior to black-box statistical approaches, the current state-of-the-art science-based approaches in the field of epidemiology have been relatively unsuccessful, in this particular instance. The reason for this is that a lot of early modeling was built by assuming some of these sophisticated models apply, but using properties analogous to other similar viruses, such as SARS or MERS, for which there was ample data available. However, this particular virus seems to behave very differently from those previous outbreaks, thus rendering those models inaccurate.

Because these efforts have proven difficult, due to the ad-hoc manner in which they must be applied, we thought it best to demonstrate our point through straightforward implementation of statistical analysis: pure regression of a causal model, with very limited assumptions. The implementation is relatively straightforward and is meant to provide an accessible explanation of just how powerful the precautionary measures of social distancing and stay-at-home orders are, and how important it is that the public obey them.

We looked at the data in two segments: before any relevant lockdown/stay-at-home measures were implemented, and after they were implemented in significant portions of the country. We observed the direct data for the period before the quarantine and modeled that as an exponential growth curve (Malthusian model) of the form:

$$f(x) = Ae^{kt} \qquad (1)$$

where $A$ is the intercept, $k$ is the exponential growth coefficient, and $t$ is time, in days. This provided us with a benchmark model for unmitigated spread of the virus. It should be noted that we did not account for any damping effect as larger portions of the population are infected. We ignored the effect of the predator-prey-like population damping, as this only becomes significant as society approaches herd-immunity levels, which are generally posited as being 60%+ of the population,[2] and we feel it would be irresponsible to even model such a scenario, since any competent society would fight tooth-and-nail to avoid this, especially because of the high CFR (Case Fatality Rate) of the disease, which is estimated at 1.4% [3], though the reported CFRs from the current data suggest a number that may be much higher. For reference, the Spanish Flu of 1918 had a CFR of ~2.5% and resulted in more than 50 million deaths [4], so the 1.4% CFR of COVID-19 is of grave concern, especially when coupled with how freely transmissible it is. As a result, in order to maintain justifiable accuracy, while avoiding indicating scenarios that should not occur in practice, we only project out the before-lockdown data a few days (well in advance of any damping effect or predator-prey behavior takes hold). The intent of projecting the no lockdown data out is to demonstrate how significant the impact of the lockdown and similar measures has been and will continue to be, and this point is sufficiently clear without the need to project out to potentially cataclysmic levels of infection.

In addition, we applied a modified exponential growth model for the after lockdown period. This was done by treating the exponent as not constant, but rather a function of time (meant to represent the impact of the lockdown). This function of time was implemented as a causal model of exponential decay. As before, the only damping force that was modeled was that of the impact of the public safety measures, as the projected levels of infection come nowhere near the level at which herd-immunity and predator-prey-like damping becomes relevant.

This can be thought of as treating social distancing and stay-at-home orders as analogous to the addition of control rods to a nuclear reaction. In the absence of the control rods, the fission continues unobstructed, as there is no shortage of particles, enabling extremely frequent collisions. However, the introduction of the control rods reduces the frequency of the collisions, because the control rods absorb the particles, reducing the number of free-moving particles that can collide. In the same way, the number of people freely moving who can spread the disease, or unto whom the disease can be spread, is greatly reduced by these social distancing and lockdown measures, which is akin to the reduction of collisions in our analogy.

Due to the noise in the data, which comes largely as a result from the data being discretized on a daily basis (and thus, a highly limited number of data points are available) and due to the high degree of variability of number of tests conducted and inconsistent testing, we chose to smooth the data as a three-day moving average of the growth rate. This is what was actually modeled for the exponential decay of the exponent.

This prediction model is constructed as follows. The general assumption of an exponential growth curve for the spread of the virus is maintained. Therefore, the following equation describes the growth is the same as equation (1), except that $k$ is no longer treated as a constant, as it was for the unmitigated growth case. Rather, $k$ is described by a causal model to account for the damping force of the public safety measures, as follows:

$$k = A_1 e^{-ct} \qquad (2)$$

where $A_1$ and $c$ are both positive valued constants that are fit to the growth rate over time, and $k$ and $t$ are the same as before.

This exponential decay function that was fit to the exponent was then applied to the data as a simple multiplicand between subsequent days and projected out, as described in the following recursion relation:

$$f_n = (f_{n-1})e^{k_n} \qquad (3)$$

where the $n$ and $n-1$ subscripts indicate the results on the $n^{th}$ and $(n-1)^{th}$ day, respectively.

As an additional note, we also only used country-wide data,[1] as the state-by-state data lacks granularity for more meaningful analysis. Further, the state-by-state data is far noisier, due to local variations in testing rates, backlogs, incomplete reporting, and inconsistent response to the coronavirus pandemic. A more macroscopic, country-wide view smooths out this data and allows for more statistically significant analysis.

It should be further understood that the delineation between the prior to lockdown and post-lockdown data is not very clear, due to inconsistent implementation across states, plus there is a lag factor due to several factors, including the time between testing and obtaining results.

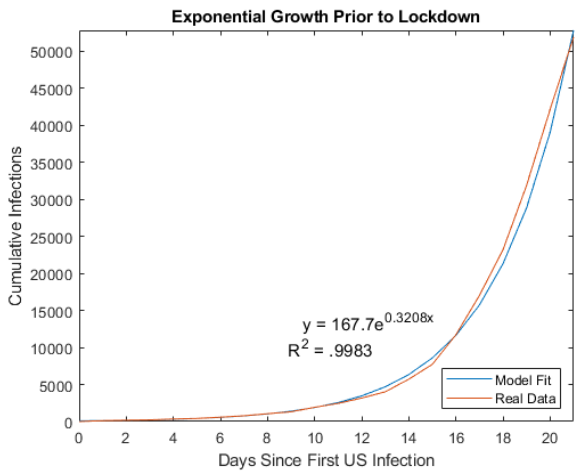

Figure 1: The exponential growth fit of the cumulative number of COVID-19 infections across the country, prior to widespread adoption of lockdown.

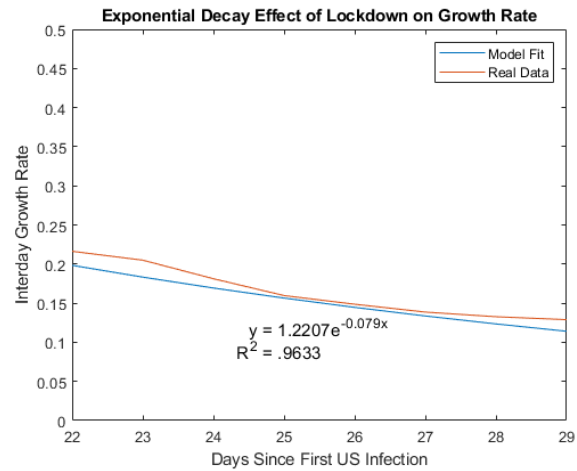

Figure 2: The causal model fit of exponential decay to inter-day growth rate of total COVID-19 infections across the country, in the days since widespread adoption of lockdown.

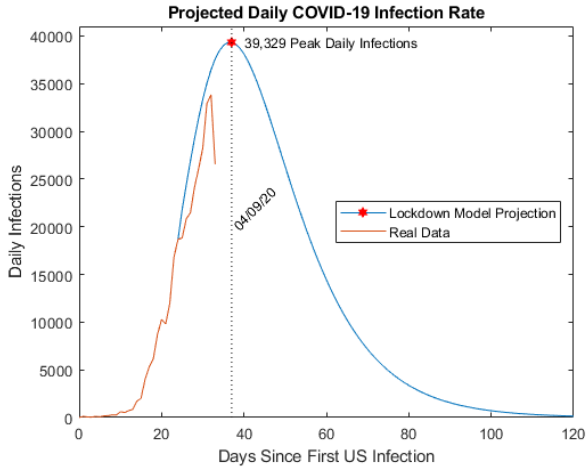

Figure 3: Daily infection rate, projected by our causal model for continued lockdown protocol, alongside real infection data.

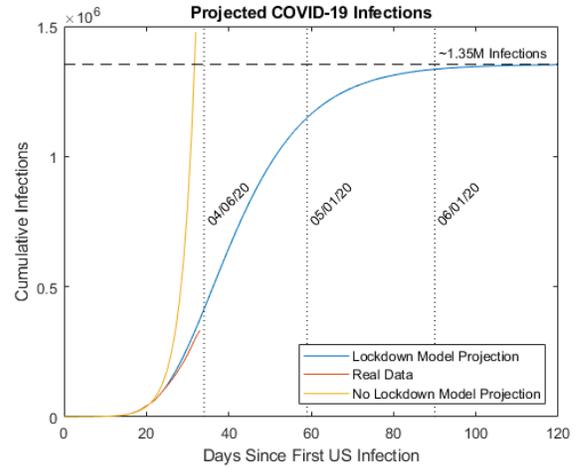

Figure 4: Cumulative infections, projected by our causal model for continued lockdown protocol, alongside real infection data, and no lockdown model projection.

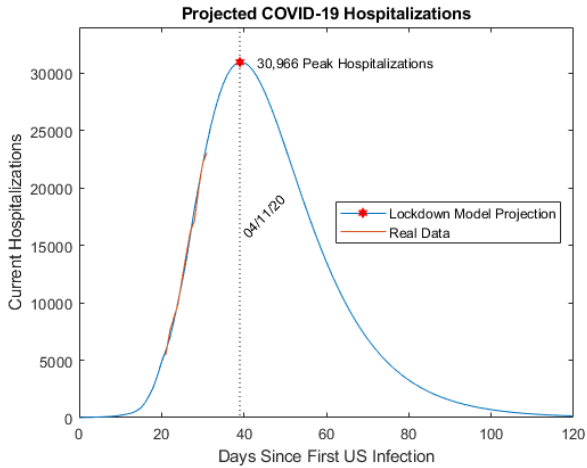

Figure 5: Active case hospitalizations, projected by our causal model for continued lockdown protocol, alongside real hospitalization data.

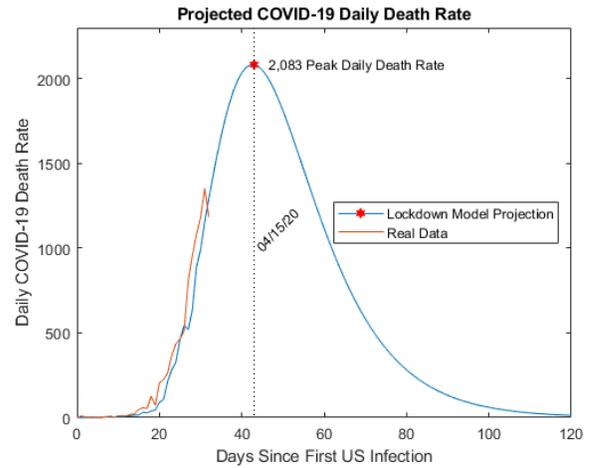

Figure 6: Daily death rate, projected by our causal model for continued lockdown protocol, alongside real COVID-19 death data.

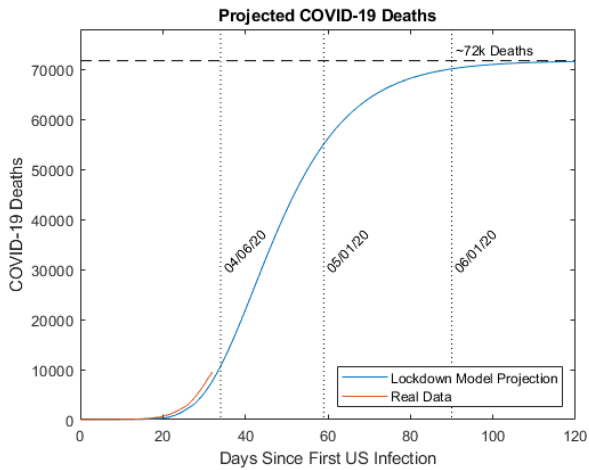

Figure 7: Cumulative deaths, projected by our causal model for continued lockdown protocol, alongside real COVID-19 death data.

### III. Results

Our analysis of the unmitigated (prior to lockdown effects) growth of infections (see Figure 1) produced the following model:

$$f(t) = 167.7e^{0.3208t} \quad (4)$$

Similarly, we acquired the following causal model for the exponential decay of the growth rate of the post-lockdown effect growth rate of infections (see Figure 2):

$$k = 1.2207e^{-0.079t} \quad (5)$$

From these models, we were able to construct figures 3 and 4, which show the daily and cumulative infections, respectively.

In Figure 3, our model indicates that daily infections, under continued lockdown protocols, should peak on April 09, 2020 (day 37) at 39,329 new infections on that day. The rate of new infections drops below 1,000 new cases per day on day 96 (June 7, 2020).

In figure 4, our model indicates that the total number of COVID-19 infections should plateau at ~1.35 Million cases, with fewer than 100 additional daily cases on day 126 (July 7, 2020).

Due to the nature of an exponential growth model with exponential decay of its exponential coefficient, there is a very long tail to continuing introduction of low levels of new cases. A real world scenario would likely see a slightly more rapid de-escalation than our model describes, in the region that is well past the peak daily infection rate.

To model the hospitalizations, we used the number of hospitalizations in the COVID tracking project to calculate a percentage of new cases requiring hospitalization, as well as calculating the mean cycle time for hospital stay (which ends in either death or recovery). From this data, we found that 13.24% of cases require hospitalization, and that the mean hospital stay is 6 days, modeled using Little's Law. Thus, we modeled the active case hospitalizations by building a recursive relationship:

$$h_i = h_{i-1} - h_{i-6} + .1324c_i \qquad (6)$$

where $h$ is the number of active case hospitalizations, the subscripts represent the day, and $c$ is the daily infection rate.

From this, we constructed figure 5, which shows that our model predicts that hospitalizations should peak on April 11, 2020 (day 39) at 30,966 hospitalizations. Should this prediction hold true, it will be a great relief, as it would be manageable by our nation's healthcare system. However, as a note, since we are looking at nationwide data, there is a possibility for inequitable distribution of these hospitalizations, such as may be the case in certain parts of New York or New Jersey, where, locally, the system may become overloaded.

From the hospitalizations, we similarly extracted the death rate. We analyzed the COVID-19 death data to determine a percentage of hospitalizations that end in death and used the same cycle time for hospitalization as determined before. As a note, there is likely a difference in terms of cycle time for recovery vs. cycle time for death, but, for lack of granularity (no knowledge of when each subset would have arrived at the hospital), we continued with the same mean hospital stay. We found that 40% of hospitalizations end in death and used the mean hospital stay of 6 days. Thus, we modeled the death rate by building a recursive relationship:

$$d_i = d_{i-1} + .4h_{i-6} \qquad (7)$$

From this, we constructed figures 6 and 7. Figure 6 shows that the death rate, according to our model, is expected to peak on April 15, 2020 (day 43) at 2083 daily deaths. This rate will drop below 200 daily deaths on day 85 (May 27, 2020) and below 100 daily deaths on day 95 (June 6, 2020).

As shown in Figure 7, our model indicates that the total death rate will plateau at ~72,000 deaths, as a result of COVID-19. While this is certainly a very high number, it is a fraction of what would occur if lockdown protocols are not continued. It should also be noted that this prediction is a far lower number than those of many existing studies, most notably the numbers put out by the CDC and White House Coronavirus Task Force [5].

## IV. Discussion and Conclusion

We do not subscribe to the notion that the prevalence of the virus is much higher, as posited by some researchers, to the level of herd-immunity. Even counting for asymptomatic cases, the number of ill people testing as positive for COVID-19 is relatively low (<20%). This is a clear, statistically significant indication that, even if the total tally is not capturing 100% of cases, we are nowhere near herd-immunity levels. Beyond rendering any analysis moot, such a notion (which has no meaningful statistical or scientific basis) is irresponsible and dangerous, as it promotes an idea that it is safe to break lockdown.

While we could attempt to draw declarative quantitative conclusions about exactly how many deaths there will be, the ventilator and ICU bed needs, the total number of cases, when we will hit our peak, and when we will see the last case of COVID-19, we choose to avoid falling for this trap. There is significant noise in the data. There is high sensitivity in the models. We have provided our models and projections and they should be thought as indicative of the number of cases, deaths, hospital requirements, and peak and fall of the pandemic, but not taken as declarative fact.

As another point, as mentioned earlier, the model predicts a very long tail to the infection.

While a long tail is generally expected for a pandemic to end, the causal model implemented, by its very nature, is a lot less aggressive in its rate of descent as the infection rate falls to low levels after the peak. A real world scenario would likely be more aggressive in the latter stages than our model indicates.

We also did not build a stochastic model and, as such, do not have error bars on our projections. The data, being discretized only on a daily basis, was too limited to construct such a model. As such, this is a limitation to our analysis.

What is very clear from this analysis, however, is that the effect of the stay-at-home and lockdown measures are profound and the absence of them, or disobeying them, is devastating.

Bootsma and Ferguson, in a paper studying the impact of the United States public health and safety measures taken in response to the outbreak of the 1918 Spanish flu in Europe, found a profound link between the speed with which such measures are implemented, how long and effectively they are maintained, and how effective they are in mitigating the spread and death rate caused by the pandemic [6]. As a result, the United States did not suffer nearly as terrible of consequences as Europe did during that pandemic.

A 60 day nationwide lockdown would most likely see us through the COVID-19 pandemic. Maintaining of general social distancing practices and use of face-masks, and improved hygiene habits (washing hands frequently, avoiding physical contact, etc.) for a period of time thereafter, upon lifting of lockdown, would allow us to transition back to a new normal, while allowing us to limit the level of the overall outbreak, as best we can. Ideally, this period of continued social distancing and general precautions would last until at least a vaccine has been developed and is readily available, or, similarly, an adequate anti-viral pharmaceutical treatment or cure is on the market and accessible to the masses. For more specificity on what that should look like, we would defer to the appropriate experts and authorities, namely the CDC and the United States Federal Government once we reach that juncture.

Acknowledgements

We would like to thank the Atlantic for having built such a useful database of nationwide and state-by-state COVID-19 data (the COVID Tracking Project) [1]. Without their freely available and accessible repository, this analysis would not have been possible. All figures were generated in MATLAB R2019a.

**Keshav Amla** is the Founder & Chief Executive Officer of Avishtech, Inc. He holds a Master of Science in Materials Science and Engineering from Stanford University and a Bachelor of Science with Honor from the California Institute of Technology.

**Tarun Amla** is the EVP & Chief Technology Officer of ITEQ Corporation and is a PhD Candidate in Mechanical Engineering at Arizona State University. He holds a Master of Science in Electrical Engineering from Stanford University, ad MBA from Northwestern University's Kellogg School of Management, a Master of Science in Mechanical Engineering (Computational Mechanics) from Purdue University, and a Master of Engineering in Modeling & Simulation from Arizona State University.

The authors declare that there is no conflict of interest.